\documentclass[12pt]{iopart}
\usepackage{amsfonts, amssymb, graphicx, float, appendix, multirow, longtable}
\usepackage{amsmath}
\usepackage{color}
\usepackage{hyperref}
\usepackage{mathrsfs}
\usepackage[ruled,vlined]{algorithm2e}
\usepackage[left=2.5cm,top=1.5cm,right=2.5cm,bottom=3cm]{geometry}
\usepackage{subfigure}

\newcommand{\sfrac}[2]{{\textstyle{#1\over#2}}}
\newcommand{\dd}{\text{D}}
\newcommand{\pp}{\partial}

\begin{document}
\title{Toward a Consistent Framework for High Order Mesh Refinement Schemes in Numerical Relativity.}

\author{Bishop Mongwane}
\address{Astrophysics Cosmology \& Gravity Center, and Department of Mathematics \& Applied Mathematics,
University of Cape Town, 7701 Rondebosch, South Africa}
\ead{astrobish@gmail.com} 

\maketitle

\begin{abstract}
It has now become customary in the field of numerical relativity to couple high order finite difference schemes to mesh refinement algorithms. To this end, different modifications to the standard Berger-Oliger adaptive mesh refinement algorithm have been proposed. In this work we present a fourth order stable mesh refinement scheme with sub-cycling in time for numerical relativity. We do not use buffer zones to deal with refinement boundaries but explicitly specify boundary data for refined grids. We argue that the incompatibility of the standard mesh refinement algorithm with higher order Runge Kutta methods is a manifestation of order reduction phenomena, caused by inconsistent application of boundary data in the refined grids. Our scheme also addresses the problem of spurious reflections that are generated when propagating waves cross mesh refinement boundaries. We introduce a transition zone on refined levels within which the phase velocity of propagating modes is allowed
to decelerate in order to smoothly match the phase velocity of coarser grids. We apply the method to test problems involving propagating waves and show a significant reduction in spurious reflections.

\end{abstract}

\section{Introduction}

Long term stable evolution of non linear hyperbolic partial differential equations often require techniques to efficiently deal with vast length scales while resolving fine scale features. Indeed, many numerical simulations in computational astrophysics, cosmology, numerical relativity and fluid dynamics are confronted with processes that span a wide range of time and length scales. 
In the context of numerical relativity, these simulations are often performed in full three spatial dimensions without symmetry assumptions. For these cases, running fine unigrid integrations is
computationally expensive and often impractical. Recent establishments in the numerical solution of partial differential equations by finite difference techniques has seen an increasing use of nested
grids and mesh refinement techniques in order to minimize the truncation error incurred with minimal computational and memory requirements \cite{1984JCoPh..53..484B,1989JCoPh..82...64B,Roma1999509}.

The principle behind mesh refinement schemes is to recursively refine areas of the computational grid that are likely to induce higher discretization errors. 
This approach efficiently focuses computational effort and resources in places where it is needed compared to refining the entire grid.
Extensive theory has since been developed for the method in different contexts, and mesh refinement algorithms have been widely adopted in the literature \cite{2000CoPhC.126..330M,2004astro.ph..3044O,Schnetter:2003rb,Debreu:2008:AAG:1316094.1316376,Ziegler2008227}. One of the key aspects in the implementation is the way in which the decision to add or remove levels of refinement is made. In this work we focus on the concept of fixed mesh refinement (FMR), where the grid hierarchy is created once and remains fixed for the duration of the computation \cite{Bruegmann:1997uc,Schnetter:2003rb,2004JCoPh.193..398C}. This differs from adaptive mesh refinement (AMR), where the algorithm is endowed with error estimation routines that dynamically determines which areas need refinement. For most applications in numerical relativity, one can know before hand which areas within the computational domain require more refinement. As a result this configuration is widely adopted in the field.

Another component of the scheme is the inter-level coupling among nested grids. While the coarse grid has to supply boundary conditions to finer grids during evolution, one can choose not to update
the coarse grid solution with the fine grid solution. This is the basis of one way (parasitic) schemes. In this work we employ the two way (interactive) scheme, where we update coarser levels with
finer levels once the finer levels have been integrated to the same time level as the coarser ones. See \cite{2010MWRv..138.2174H} for a comparison between parasitic and interactive coupling.

Traditionally mesh refinement techniques were coupled to second order convergent methods. On the other hand, recent trends in the numerical simulation community has seen the coupling of higher order finite difference methods to the mesh refinement framework \cite{Shen:2011:AMR:1963655.1963728,Lehner:2005vc}. This combines the efficiency of local mesh refinement with the robustness and accuracy of higher order methods. 
 However, there is an inherent incompatibility between high order time discretization schemes with the standard mesh refinement algorithm that may result in loss of convergence or even instabilities \cite{Lehner:2005vc}. This issue is related to how the computation of boundary data for the refined grids is handled. 
A search for a stable high order mesh refinement implementation has resulted in several modifications to the standard method in an effort to address this subject. Most notably, \cite{Schnetter:2003rb}
introduces the idea of buffer zones in the refined grids. In this setup, boundary conditions are not prescribed explicitly in the refined levels, the integration is only applied to a progressively
smaller domain in the refined grids and the buffer zone is ultimately discarded. Another approach is the tapered boundary approach \cite{Lehner:2005vc,Csizmadia:2007zz}. Here, one performs the
integration at level $l$ using the past domain of dependence of the child grid only. Other approaches have been to refine only in space and using the same time step for all levels
\cite{2004JCoPh.193..398C}. In this work we use a framework where we refine both in space and time and the treatment of interface boundaries is dictated by the time marching algorithm, fourth order
accurate Runge Kutta algorithm in this case.

In addition to issues of convergence and stability, one has to address the problem of spurious reflections off refinement boundaries that arise when waves cross refinement boundaries. This is
essential for gravitational wave source simulations as the waves are normally extracted at a large radius. Propagating waves will have crossed several refinement boundaries, before reaching the radius
of extraction. In \cite{2004JCoPh.193..398C}, the idea of derivative matching was proposed in order to minimize spurious reflections for second order convergent schemes. Also, the concept of mesh
adapted stencils (MAD) was introduced in \cite{Baker:2005xe}. However these implementations do not involve refinement in time. Other methods that have been applied in the Advanced Weather Research and
Forecasting Model, Euler equations and Maxwell equations involve the use of sponge layers in the refined levels to ensure that the solution in the refined levels will be nudged towards that of the
coarser grids at refinement boundaries. This may involve the addition of artificial damping and dissipation terms in the system under consideration \cite{2010MWRv..138.2174H,Skamarock01adescription}.
See also the treatment of \cite{MohanRai1986472} in the case of first order convergent schemes. In this work we propose a simple scheme that is adopted from the animation and image processing
community \cite{ebert02texturing}, to deal with transitions from fine to coarse grid solutions at refinement boundaries.

This paper is organized as follows: in Section \ref{sec:generalities} we convey the framework of our FMR approach. We review our boundary application method in \S\ref{sec:ghostzones} and introduce the
transition zone in \S\ref{sec:transition_zone}. We present our results in Section \ref{sec:numerical_examples} and finally concluding in Section \ref{sec:concluding_remarks}.

\section{Generalities}
\label{sec:generalities}

\subsection{Grid layout}

\begin{figure}[!h]
\centering
\includegraphics[width=7.5cm,height=6cm]{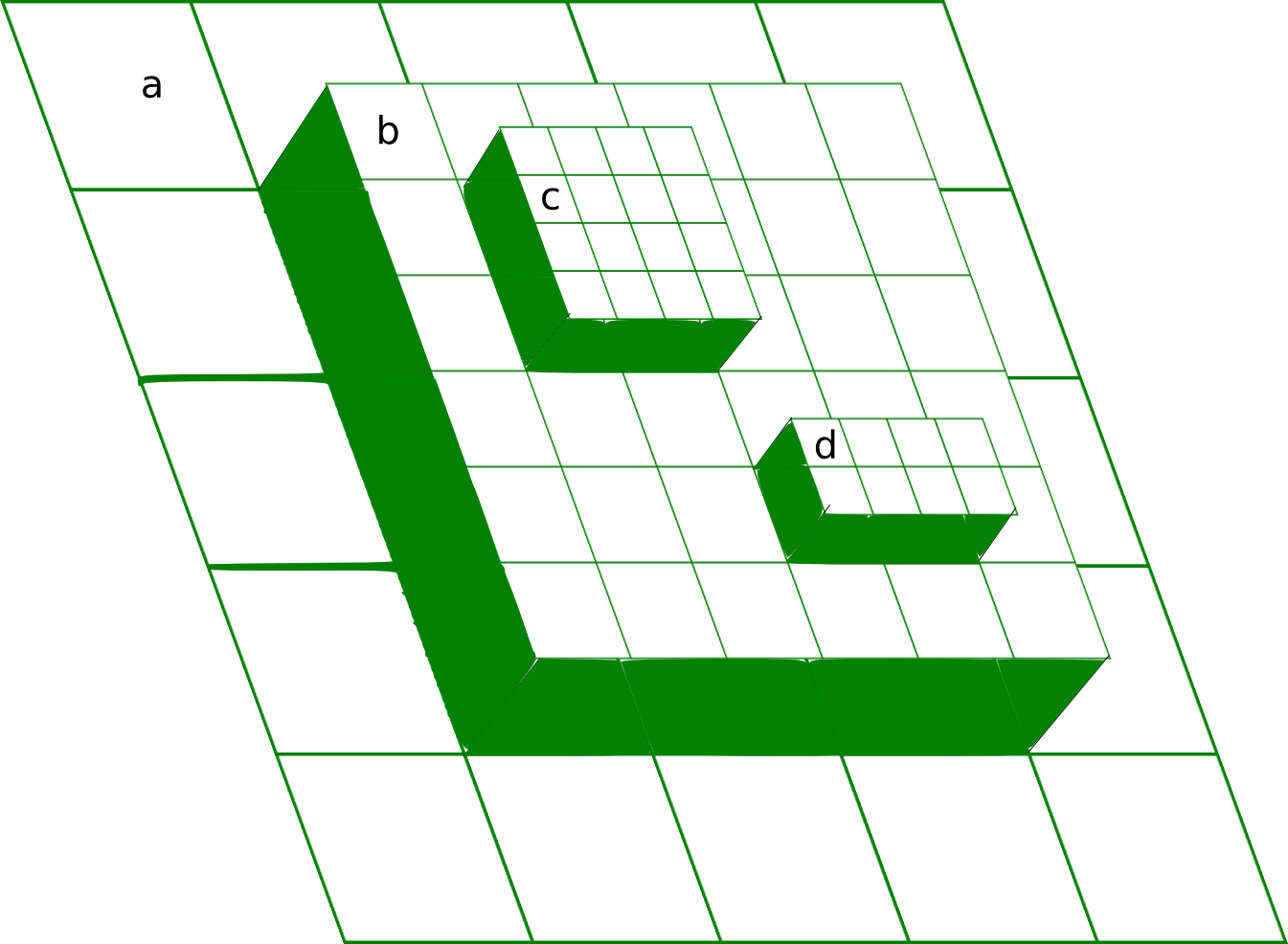}
\caption{\small{\textit{A grid hierarchy demonstrating proper nesting. A single grid $ H^{0}_{~~0}$ covering the entire domain is marked by `a'. There is one refinement grid $H^{1}_{~~0}$ at level one
marked by 'b'. Two disjoint refinement grids are $H^{2}_{~~0}$ and $H^{2}_{~~1}$ are marked by `c' and `d' respectively. Note that the ghost zones are not included in the grids. This figure is used to
emphasize that the grid hierarchy is not some complex data structure, but that the overlaid grids are independently stored in memory.}}}
\label{fig:grid_hierachy2}
\end{figure}

The grid hierarchy is arranged by first discretizing the spatial domain into a relatively coarse uniform mesh that covers the entire computational domain.
This constitutes the base or root grid $H^{0}_{\phantom{0} 0}$ with mesh size $h_{0}$. Finer grid patches of mesh sizes $h_{l}/r$ are then overlaid as required to the base grid with each grid at level
$l$ having mesh size $h_{l-1}/r$. More than one grid patch can be added in a given level. This forms a tree or a hierarchy of grids $H^{l}_{\phantom{p}p}$, where the indices $l$ and $p$ represent the
level and patch number respectively. This configuration is depicted in Figure \ref{fig:grid_hierachy2}. It is at this point that we emphasize that each grid in a given level $l$ has its own solution
vector and is evolved independently of all other grids. Of course it has to depend on the parent grid, within which it is nested, for boundary data. 

Each grid patch added at a given level must satisfy certain conditions. Among them, the idea of proper nesting: A fine grid at level $l$ must start and end at the corner of a cell belonging to level $l-1$. Moreover, grids at higher levels cannot `float'. This means that if there is a grid at level $l+2$, it must be contained in a grid at level $l+1$ that is itself properly nested on a grid at level $l$.
The refinement factor $r$ must be an integer, and is the same for all levels. This results in a constant CFL for all added levels, thus the same integration routine of the base level is stable on all levels, if it is stable on level $0$. It also implies that grids at higher levels require $r^{l}$ time steps to catch up with a single time step of the base grid.


\subsection{Inter-level communication}
Each grid in a given level $l$ can be indexed independently by its own $(i_{l}, j_{l}, k_{l})$ coordinate system. However, for reasons of inter-level communications, there is a mapping from the
$H^{l}_{\phantom{p}p}$ coordinates $(i_{l}, j_{l}, k_{l})$ to the $H^{l-1}_{\phantom{p}p}$ coordinates $(i_{l-1}, j_{l-1}, k_{l-1})$ and vice-versa. This relation is expressed as,
\begin{equation}
\label{eq:index_map}
i_{l-1} = \frac{i_{l} - \mathbf{mod}(i_{l},r)}{r}
\end{equation}
for a staggered hierarchy. Such communications are necessary for the computation of initial conditions, boundary conditions and for updating the coarse grid with the fine grid solution.
Initial data can be generated by spatial interpolation from the previous grid level (\textit{prolongation}) or by calling the same initialization routine that was used to initialize the base grid. All
levels are added and initialized at the same initial time $t_{0}$. 
%
Once all levels have been integrated to the same time, data in the finer meshes is used to update data in the coarse levels through the use of interpolation operators, a process called \textit{restriction}.

It is important to note that with the mapping of indices \ref{eq:index_map}, all fine grid points are staggered about coarse grid points at lower levels. This has the advantage that if the base grid
is discretized strategically to `avoid' certain points $(x, y, z) \in \mathbb{R}^{3}$, (e.g. by using a cell centered grid to avoid dealing explicitly with the points on the edges) such points remain
avoided in all refined levels. However, from a computational standpoint, this may be expensive because communicating data across levels requires three dimensional interpolation all the time. This is
different from cases where some points in the fine grids are allowed to coincide with the ones from coarser levels. For such cases, inter-level communication can occur via \textit{injection}, where
data is simply copied to corresponding points at a given level.

\subsection{Numerical Integration}

Spatial differentiation is handled through fourth order convergent finite differencing. The first derivative is given by the operator,
\begin{align}
\label{eq:firstDeriv}
  \pp_x f_{i,j,k} = \frac{f_{i-2,j,k} - 8 f_{i-1,j,k} + 8 f_{i+1, j,k} - f_{i+2,j,k}}{12 dx},
\end{align}
while the second derivative is given by,
\begin{align}
  \pp_{xx} f_{i,j,k} = \frac{- f_{i+2,j,k} + 16 f_{i+1,j,k} -30 f_{i,j,k} + 16 f_{i-1,j,k} - f_{i-2,j,k}}{12 dx^2}.
\end{align}
The cross derivative is given by sequential application of \ref{eq:firstDeriv}. In the case of the $xy$ derivative, this results in,
\begin{align}
  \pp_{xy} f_{i,j,k} = \frac{\pp_{y}f_{i-2,j,k} - 8 \pp_{y}f_{i-1,j,k} + 8 \pp_{y}f_{i+1, j,k} - \pp_{y}f_{i+2,j,k}}{12 dx},
\end{align}
We use these centered stencils for all derivatives, except advection terms for which we use the lop sided formulas,

\begin{equation}
  \pp_x f_{i,j,k} = \frac{-f_{i-3,j,k} + 6 f_{i-2,j,k} - 18 f_{i-1, j,k} + 10f_{i,j,k} + 3f_{i+1,j,k}}{12 dx}, \qquad \beta^{x} < 0
\end{equation}

\begin{equation}
  \pp_x f_{i,j,k} = \frac{f_{i+3,j,k} - 6 f_{i+2,j,k} + 18 f_{i+1, j,k} - 10f_{i,j,k} - 3f_{i-1,j,k}}{12 dx}, \qquad \beta^{x} > 0
\end{equation}

Time integration is carried out using a fourth order accurate Runge Kutta scheme through the method of lines framework. A Runge Kutta step from $y_{n}$ to $y_{n+1}$ is accomplished with,
\begin{equation}
\label{eq:step}
y_{n+1} = y_{n} + \frac{1}{6} (k_{1} + 2k_{2} + 2k_{3} + k_{4})\,,
\end{equation}

\begin{equation}
\label{eq:classicform}
\begin{aligned}
  k_{1} &= h f\left( x_{n}, Y_{1} \right) \\
  k_{2} &= h f\left( x_{n} + \sfrac{1}{2}h, Y_{2} \right)\\
  k_{3} &= h f\left( x_{n} + \sfrac{1}{2}h, Y_{3} \right) \\
  k_{4} &= h f\left( x_{n} + h, Y_{4} \right) \\
\end{aligned}
\end{equation}
where the quantities $Y_{i}$ defined by,
\begin{equation}
\label{eq:apply_boundary_conditions}
\begin{aligned}
  Y_{1} = y_{n}, \qquad  Y_{2} = y_{n} + \sfrac{1}{2}k_{1},  \qquad Y_{3} = y_{n} + \sfrac{1}{2}k_{2}, \qquad\mathrm{and} \qquad Y_{4} = y_{n} + k_{3} .
\end{aligned}
\end{equation}
serve to store boundary data.

The control algorithm that is responsible for the evolution of the entire grid hierarchy is orchestrated by a recursive procedure, shown in Algorithm \ref{alg:integrate_hierachy}

\bigskip
\begin{algorithm}[!h]
\label{alg:integrate_hierachy}
\SetAlgoVlined

Procedure propagate()\\ 
\KwIn{\textbf{int} level}
\KwOut{void}
advanceLevel(level) \;

\uIf{$level < max\_level$}
    {
      \ForEach{iteration $j = 1$, r}
              {
                propagate(level + 1) \;
              } 

    }
\Return \;
\caption{\small{\textit{A simple illustration of the AMR integration algorithm with refinement factor $r$.}}}
\end{algorithm}
\bigskip

\subsection{Boundary Treatment}
The base grid consist of a ghost zone that serve to store boundary data. For refined grids, in addition to ghost zones, there is also a transition zone that ensures a smooth transition from the well resolved solution of the finer grids to the less resolved solution of the coarser grids. We discuss each in turn.

\subsubsection{Ghost zone}
\label{sec:ghostzones}
For a fourth order accurate computation of centered derivatives, we use two ghost zones on each side of a grid\footnote{However, our implementation is such that one needs the number of Ghost points to
be odd for a staggered mesh and even otherwise. We will employ the staggered mesh in this paper and thus choose the number of Ghost points to be three.}.
Outer boundary conditions on the base grid are supplied by the user, for example, the user may specify periodic or sommerfeld type boundary conditions.
We distinguish between two types of boundary conditions for the finer levels, those that coincide with the outer boundary and those that simply border a cell from the underlying coarse grid.

If the ghost zones of refined grids coincide with those of the base grid, the boundary is filled using the prescribed procedure for outer boundaries. Otherwise, we use coarse grid data to fill the
fine grid boundaries. In \cite{bish:2014}, we showed that using the conventional method of imposing boundary conditions, i.e, simply applying boundaries corresponding to the intermediate times of
Runge Kutta methods leads to a loss of convergence for unigrid runs. We have found that this method, leads to unstable modes in the mesh refinement case. We instead use the Runge Kutta method itself
to fill the ghost zones. We use Equations \ref{eq:apply_boundary_conditions}, where the $k_{i}$ are given by, \cite{bish:2014}
%
%
%
\begin{equation}
\label{eq:non_linear_scheme}
\begin{aligned}
k_{1} & =  hy'\\
k_{2} & =  hy' + \frac{h^{2}}{2}y'' + \frac{h^{3}}{8}\left(y''' - f_{y}y''\right)  \\
k_{3} & =  hy' + \frac{h^{2}}{2}y'' + \frac{h^{3}}{8}\left(y'''+ f_{y}y''\right) 
\end{aligned}
\end{equation}
In the equations above, $y', y''$ and  $y'''$ are time derivatives of the quantities under evolution while $f_{y}$ is the Jacobian matrix of the PDE system. See \cite{bish:2014} for more details. The time derivatives can easily be obtained by polynomial interpolation methods since the coarse grid points at the advanced time will already have been computed before advancing the refined levels.
However, a subtle issue arises in this case. To evolve the finer grids, at least four past points of the coarse grid solution are needed in order to obtain third order interpolants. This means each
finer grid can only be initialed after the coarser grid has evolved at least four time steps. This is undesirable in the context of FMR. We opt to use the fact that the classical Runge Kutta method
has a built-in interpolant, termed dense output. This interpolant is given by, \cite{Hairer2009Book}
\begin{equation}
\label{eq:interpolant}
y(t_{n}+\theta h) = y_{n} + \sum_{i=1}^{4} b_{i}(\theta)k_{i} + \mathcal{O}(h^{4})
\end{equation}
with $0 \leq \theta \leq 1$ and the $b_{i}$ are polynomials in $\theta$,
\begin{equation}
b_{i}(\theta) = \theta - \frac{3}{2}\theta^{2}+\frac{2}{3}\theta^{3}, \qquad  b_{2}(\theta) = b_{3}(\theta) = \theta^{2} - \frac{2}{3}\theta^{3}, \qquad \mathrm{and} \qquad b_{4}(\theta) = -\frac{1}{2}\theta^{2}+\frac{2}{3}\theta^{3}
\end{equation}
One can verify that this dense output formula reduces to Equation \ref{eq:step} when $\theta=1$. The required time derivatives are then computed from Equation \ref{eq:interpolant} as,
\begin{eqnarray}
\frac{d^{(m)}}{dt^{(m)}}y(t_{n}+\theta h) &=& \frac{1}{h^{m}}\sum_{i=1}^{s} k_{i} \frac{d^{(m)}}{d\theta^{(m)}}b_{i}(\theta) + \mathcal{O}(h^{4-m})
\end{eqnarray}
with $1\leq m \leq 3$. One does not need to compute the Jacobian matrix $f_{y}$ explicitly since we are only interested in the product $f_{y}y''$ which can be computed from the system \ref{eq:non_linear_scheme} as,
\begin{equation}
f_{y}y'' = \frac{4}{h^{3}}\left(k_{3}-k_{2}\right)
\end{equation}
See also, \cite{mccorquodale2011high}. The implication is that we do not store the solution history of coarser grids but we store the four (current) intermediate $k_{i}$ values instead. Of course this
is followed by spatial interpolation, for which we employ fourth order barycentric Lagrange interpolation; higher than fourth order was found to be unreliable in some of the runs.

\subsubsection{Transition zone}
\label{sec:transition_zone}
\begin{figure}[!h]
\centering
\includegraphics[height=7cm,width=0.6\textwidth]{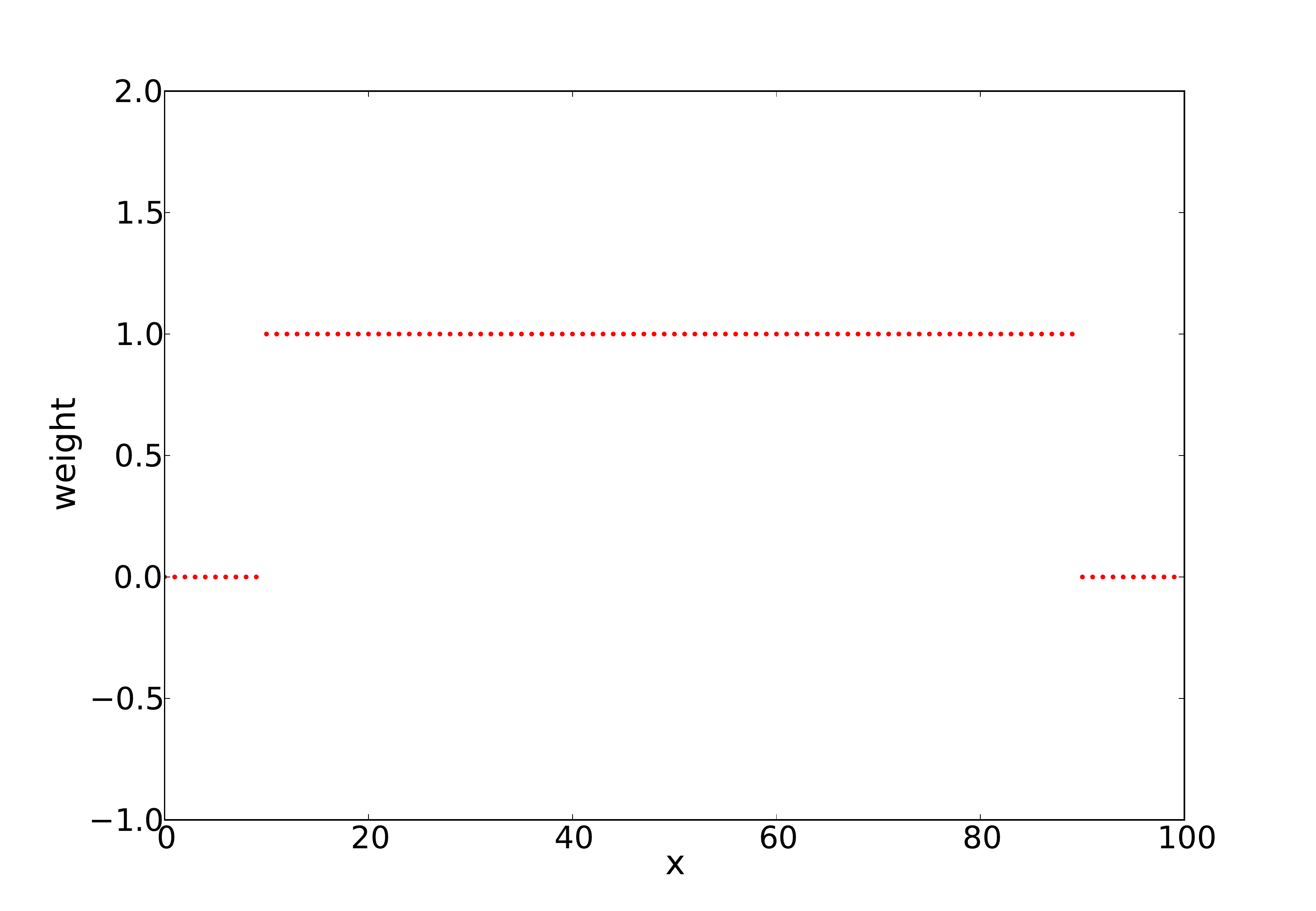}
\caption{\small{\textit{Step function transition profile from $w=0$ to $w=1$. The refined region in this case is $x\in[10,90]$. Note the discontinuity at $x=10$ where the weight $w$ transitions from w=0 to w=1 and again at $x=90$ where $w$ transitions from $w=1$ to $w=0$.}}}
\label{fig:step}
\end{figure}

To complete the specifications on treatment of the boundary, we examine what happens close to the refinement boundary. 
Consider a grid hierarchy with two levels $l_{0}$ and $l_{1}$. Parametrize the solution $F(x)$ on such a hierarchy as,
\begin{equation}
\label{eq:transition}
F(x) = (1-w)f(x,l_{0}) + wf(x, l_{1}),
\end{equation}
where $f(x,l_{0})$ and $f(x,l_{1})$ are the solutions on the base and refined grids respectively, and $w$ is a binary weight function which takes the value $w=1$ if $x$ is within the refined region
and $w=0$ everywhere else. A plot of the weight function is depicted in Figure \ref{fig:step}. Note the discontinuity at the transition points $x=10$ where $w$ transitions from $w=0$ to $w=1$
indicating a switch from the solution $F(x)=f(x,l_{0})$ to $F(x)=f(x,l_{1})$. Also at $x=90$, $w$ transitions from $w=1$ to $w=0$, indicating the switch from $F(x)=f(x,l_{1})$ back to $F(x) =
f(x,l_{0})$.

\begin{figure}[!h]
\centering
\includegraphics[height=7cm,width=0.6\textwidth]{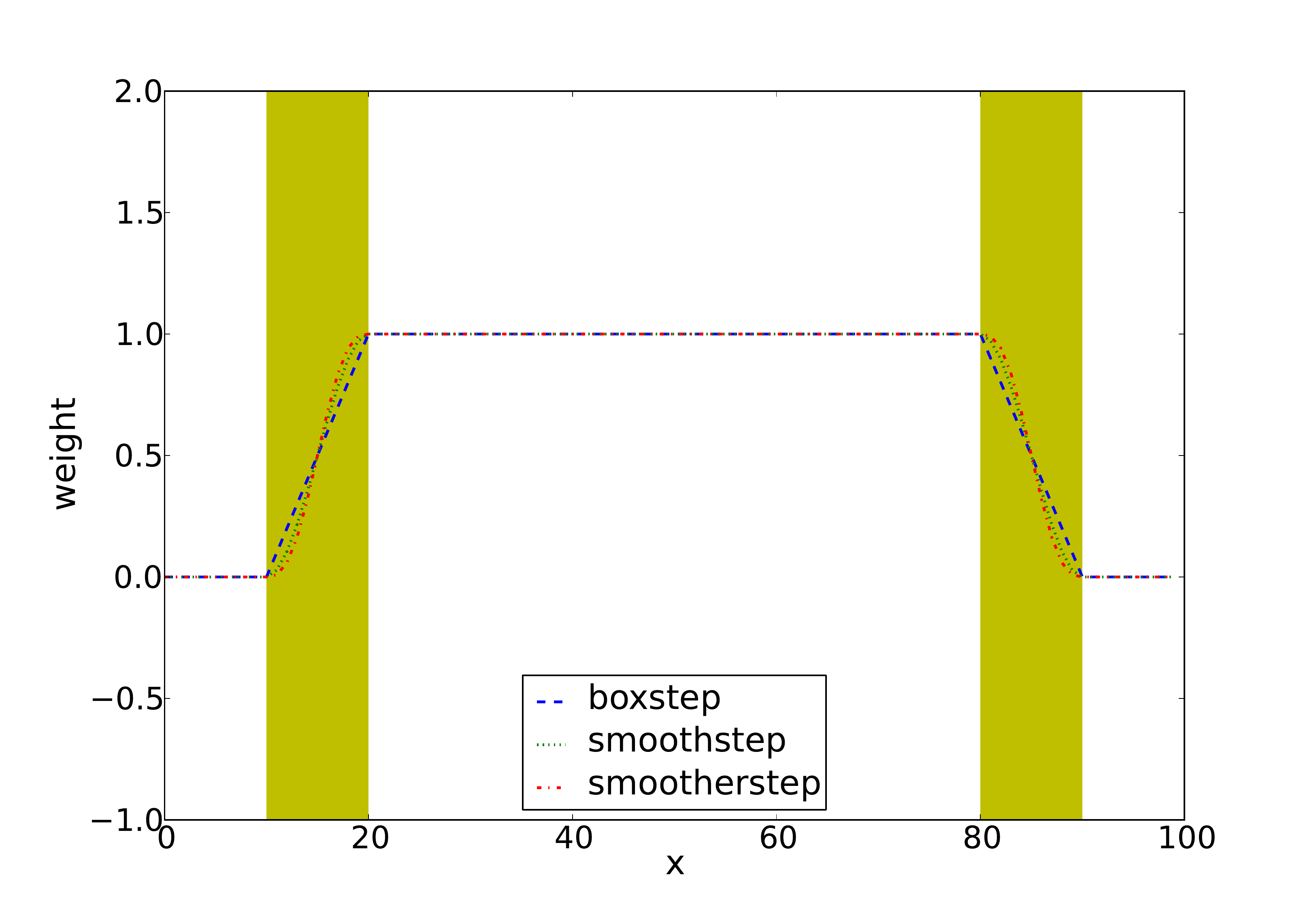}
\caption{\small{\textit{Smooth transition profiles from $w=0$ to $w=1$. The refined region in this case is $x\in[10,90]$ with the shaded regions representing the transition zone. We have exaggerated
the width of the transition zone for ease of visualization. Compare with Figure \ref{fig:step}.}}}
\label{fig:profiles}
\end{figure}

Because of the dispersion relation for propagating waves (\S \ref{sec:dispersion}), there is a difference in phase speeds of propagating modes in the coarse and fine grid levels. As a result of the discontinuous transition in the weight function $w$, waves propagating from refined regions abruptly change their phase velocities when crossing refinement boundaries, creating a glitch that will seed spurious reflections.
To circumvent this problem, we introduce a transition zone on the refined levels, within which the weight function $w=w(x)$ is allowed to vary smoothly from $w=0$ to $w=1$ across the refinement
boundary. This can be accomplished by Hermite interpolation. For a transition beginning at $x=a$ and ending at $x=b$, one can derive the following profiles,
\begin{align}
\label{eq:profiles}
w(a,b,x) &= t \qquad&  \mathrm{(boxstep)}\\
w(a,b,x) &= 3t^{2}-2t^{3}\qquad& \mathrm{(smoothstep)}\\
w(a,b,x) &= 10t^{3}-15t^{4}+6t^{5} \qquad& \mathrm{(smootherstep)}
\end{align}
where the variable $t$ is defined as,
\[ t=\begin{cases} 
      0 & \frac{x-a}{b-a} < 0 \\
      1 & \frac{x-a}{b-a} > 1 \\
      \frac{x-a}{b-a} & \mathrm{otherwise}
   \end{cases}
\]
These profiles are shown in Figure \ref{fig:profiles}. In this case the weight function varies continuously from $w=0$ to $w=1$, allowing the solution $F(x)$ to vary smoothly across the transition zone. This can also be interpreted as a smooth acceleration and deceleration of the associated phase speeds of propagating modes in the solution $F(x)$. See \cite{ebert02texturing} for a discussion on transition profiles. 

The procedure we follow to ensure a smooth transition from the fine grid solution to the coarse grid solution is as follows.
\begin{itemize}
\item Fill ghost zone points using the method outlined in \S\ref{sec:ghostzones}.
\item Fill the transition zone by blending values from the refined grid, with that of the coarse grid according to the weight function $w(x)$.
\end{itemize}
The transition zone is evolved along with the fine grid solution to ensure a smooth coupling with the refinement boundary and thus the coarse grid solution. However, for reasons of stability, we do not use transition zone values when updating the coarse grid solution with the fine grid solution. Unless otherwise specified, we take the width of the transition zone to be three through out this work. For this size, the smooth profiles given above are equivalent.

\section{Evolution system}

We adopt the BSSN formulation of the Einstein field equations \cite{Shibata:1999yx,Baumgarte:1998te}. The evolution equations are given in terms of the variables,
\begin{eqnarray}
\tilde \gamma_{ij} &=& e^{- 4 \phi} \gamma_{ij} \\
\tilde{A}_{ij} &=& e^{- 4 \phi}(K_{ij} - \frac{1}{3} \gamma_{ij} K)
\end{eqnarray}
subject to the constraints, $\tilde{\gamma}=\mathrm{Det}\;\gamma_{ij}=1$, $\tilde{A}^{i}_{\phantom{i}j}=0$. Additional variables $\tilde{\Gamma}^{i}=- \tilde{\gamma}^{ij}_{\phantom{i},j}$ are also introduced. The evolution equations for these variables are derived from the ADM equations and are given by,


\begin{eqnarray}
\pp_{t} \phi & = & -\frac{1}{6}\alpha K + \beta^{k}\pp_{k}\phi + \frac{1}{6}\pp_{k}\beta^{k},\label{eq:phidot}\\
\pp_{t} \tilde{\gamma}_{ij} & = & -2\alpha\tilde{A}_{ij} + \beta^{k}\pp_{k} \tilde{\gamma}_{ij} + \tilde{\gamma}_{ik}\pp_{j}\beta^{k} + \tilde{\gamma}_{jk}\pp_{i}\beta^{k} - \frac{2}{3}\tilde{\gamma}_{ij}\pp_{k}\beta^{k},\label{eq:gdot}\\
\pp_{t} K & = &
	\alpha\left(\tilde{A}_{ij}\tilde{A}^{ij}+\frac{1}{3}K^2
	\right)-\gamma^{ij}\dd_{i}\dd_{j}\alpha + \beta^{k}\pp_{k}K \label{eq:Kdot}\\
\pp_{t} \tilde{A}_{ij} & = &
	\alpha\left(K\tilde{A}_{ij}-2\tilde{A}_{ik} 
   	\tilde{A}^k{}_j\right)
   	+e^{-4\phi}\big(\alpha R_{ij} -\dd_{i}\dd_{j}\alpha\big)^{TF} + \nonumber \\
        &&\beta^{k}\pp_{k} \tilde{A}_{ij} + \tilde{A}_{ik}\pp_{j}\beta^{k} + \tilde{A}_{jk}\pp_{i}\beta^{k} - \frac{2}{3}\tilde{A}_{ij}\pp_{k}\beta^{k},
   	\label{eq:Adot}\\
\pp_{t}\tilde{\Gamma}^i & = & 2\alpha\left(\tilde{\Gamma}^i_{jk}
    \tilde A^{jk}-\frac{2}{3}\tilde{\gamma}^{ij}K_{,j}
    +6\tilde{A}^{ij}\phi_{,j}\right) 
    -2\tilde{A}^{ij}\alpha_{,j}+\tilde{\gamma}^{jk}\beta^i{}_{,jk}+
	\frac{1}{3}
    \tilde{\gamma}^{ij}\beta^k{}_{,jk}+\beta^j\tilde{\Gamma}^i{}_{,j}
    \nonumber\\
   &&-\tilde{\Gamma}^j\beta^i{}_{,j} 
    +\frac{2}{3} \tilde{\Gamma}^i\beta^j{}_{,j}.\label{eq:Gammaidot}
\end{eqnarray}
The superscript $TF$ denotes the trace-free part with respect to the metric $\gamma_{ij}$, and
\begin{equation}
\dd_{i}\dd_{j}\alpha = \pp_{i}\pp_{j}\alpha - 4\pp_{(i}\phi\pp_{j)}\alpha-\tilde{\Gamma}^{k}_{ij}\pp_{k}\alpha+
2\tilde{\gamma}_{ij}\tilde{\gamma}^{kl}\pp_{k}\phi \pp_{l}\alpha.
\end{equation}
The Ricci tensor $R_{ij}$ is now written as a sum of two pieces
\begin{equation}
   R_{ij} = \tilde{R}_{ij} + R^{\phi}_{ij},
\end{equation}
where $R^{\phi}_{ij}$ is given by

\begin{eqnarray}
   R^{\phi}_{ij}&=&-2\tilde{\dd}_i\tilde{\dd}_j\phi-2\tilde{\gamma}_{ij}
   \tilde{\dd}^k\tilde{\dd}_k\phi + 4\tilde{\dd}_i\phi\tilde{\dd}_j\phi-4\tilde{\gamma}_{ij}\tilde{\dd}^l\phi
   \tilde{\dd}_l\phi, \\
   \tilde{R}_{ij}&=&-\frac{1}{2}\tilde{\gamma}^{mn}\tilde{\gamma}_{ij,mn}
     +\tilde{\gamma}_{k(i}\tilde{\Gamma}^k{}_{,j)}
     +\tilde{\Gamma}^k\tilde{\Gamma}_{(ij)k} 
     +\tilde{\gamma}^{mn}\left(2\tilde{\Gamma}^k{}_{m(i}\tilde{\Gamma}_{j)kn}
     +\tilde{\Gamma}^k{}_{in}\tilde{\Gamma}_{kmj}\right).
\end{eqnarray}



\section{Numerical results}
\label{sec:numerical_examples}
For ease of exposition, we restrict to the case $\beta^{i}=0$.
The time step is given by $dt=c dx$ where $c=0.25$ for all the runs considered here. Where refinement is used, we restrict to a refinement factor $r=2$. All runs employ three ghost points and, where appropriate, three transition points.

\subsection{Wave equation: Periodic boundaries}
\begin{figure}[h!]
\centering
\includegraphics[height=7.5cm, width=0.7\textwidth]{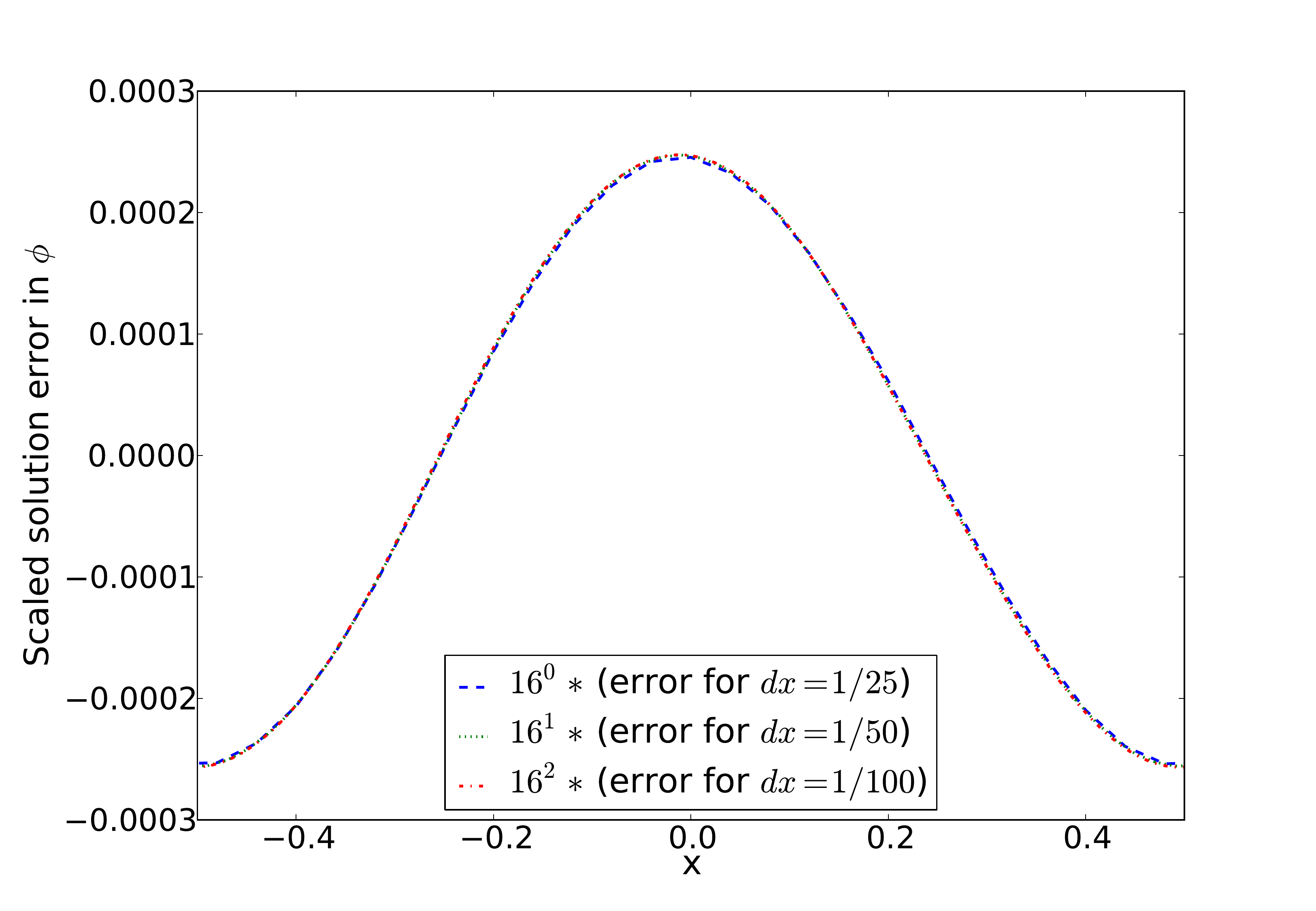}
\caption{\small{\textit{Scaled solution errors for $\phi$ after $2$ crossing times. The errors have been scaled with the resolution to highlight fourth order convergence.}}}
\label{fig:converge_wave}
\end{figure}

In this section, we carry out evolutions of the wave equation with mesh refinement. The wave equation in flat Cartesian coordinates is given by,
\begin{equation}
\pp_{tt} \phi = \pp^{i} \pp_{i} \phi.
\end{equation}
We instead cast it in an alternative form, by introducing a new auxiliary variable $\Pi=\pp_{t}\phi$,
\begin{align}
\label{eq:wave_eq2}
\pp_{t} \phi &= \Pi\\
\pp_{t} \Pi &=  \pp^{i} \pp_{i} \phi 
\end{align}
As initial data, we choose a sinusoidal profile
\begin{equation}
\phi(x,y,z, t=0) = \sin(2\pi(x-t)) \qquad \Pi(x,y,z,t=0) = -2\pi\cos(2\pi(x-t)),
\end{equation}
 with periodic boundary conditions on the domain $x\in[-0.5,0.5]$. The region $x\in[-0.25,0.25]$ is refined by a factor of $r=2$. Although the wave propagates essentially in one dimension, we evolve it using the full 3D grid with periodic boundary conditions for the outer boundaries. In Figure \ref{fig:converge_wave} we plot the solution errors for evolutions with resolution $dx=1/25\rho$ for $\rho=1,2,3$. The errors show fourth order convergence as desired.

\subsection{Gauge Wave}
\label{sec:gauge_wave}
We now evolve a non linear gauge wave using the BSSN system.
The Gauge wave test is characterized by a line element which results from a non linear gauge transformation of the flat Minkowski space time in Cartesian coordinates, resulting in
\begin{equation}
ds^{2} = -H dt^{2}+ Hdx^{2}+dy^{2} + dz^{2}\;,
\end{equation}
where the function $H$ is given as,
\begin{equation}
H\equiv H(x-t) = 1 + A \sin\left(\frac{2\pi(x - t)}{d} \right)\;,
\end{equation}
for some constant $A$, and $d$ is the wavelength. We evolve the above metric using the BSSN formulation with the Harmonic gauge condition,
\begin{equation}
\pp_{t} = -\alpha^{2}K\;.
\end{equation}
We choose the amplitude $A=0.1$ and the wavelength $d=1$. As in the last case, the simulation domain covers the range $x\in[-0.5:0.5]$ with refinement boundaries in the region $x\in[-0.25:0.25]$. Evolving the gauge wave initial data with the BSSN formulation requires the addition of artificial dissipation to achieve stable evolutions. This is true even for unigrid runs, see for example \cite{Babiuc:2007vr}. Our dissipation operator takes the form
\begin{equation}
\label{eq:ko_dissip}
\pp_{t}Q \rightarrow \pp_{t}Q+(-1)^{r/2}\sigma \sum_{i}h_{i}^{r+1}\mathcal{D}_{i+}^{r/2+1} \mathcal{D}_{i-}^{r/2+1}Q\;,
\end{equation}
for r$th$ order accurate finite difference stencils. 

\begin{figure}[h!]
\centering
\includegraphics[height=7.5cm, width=0.7\textwidth]{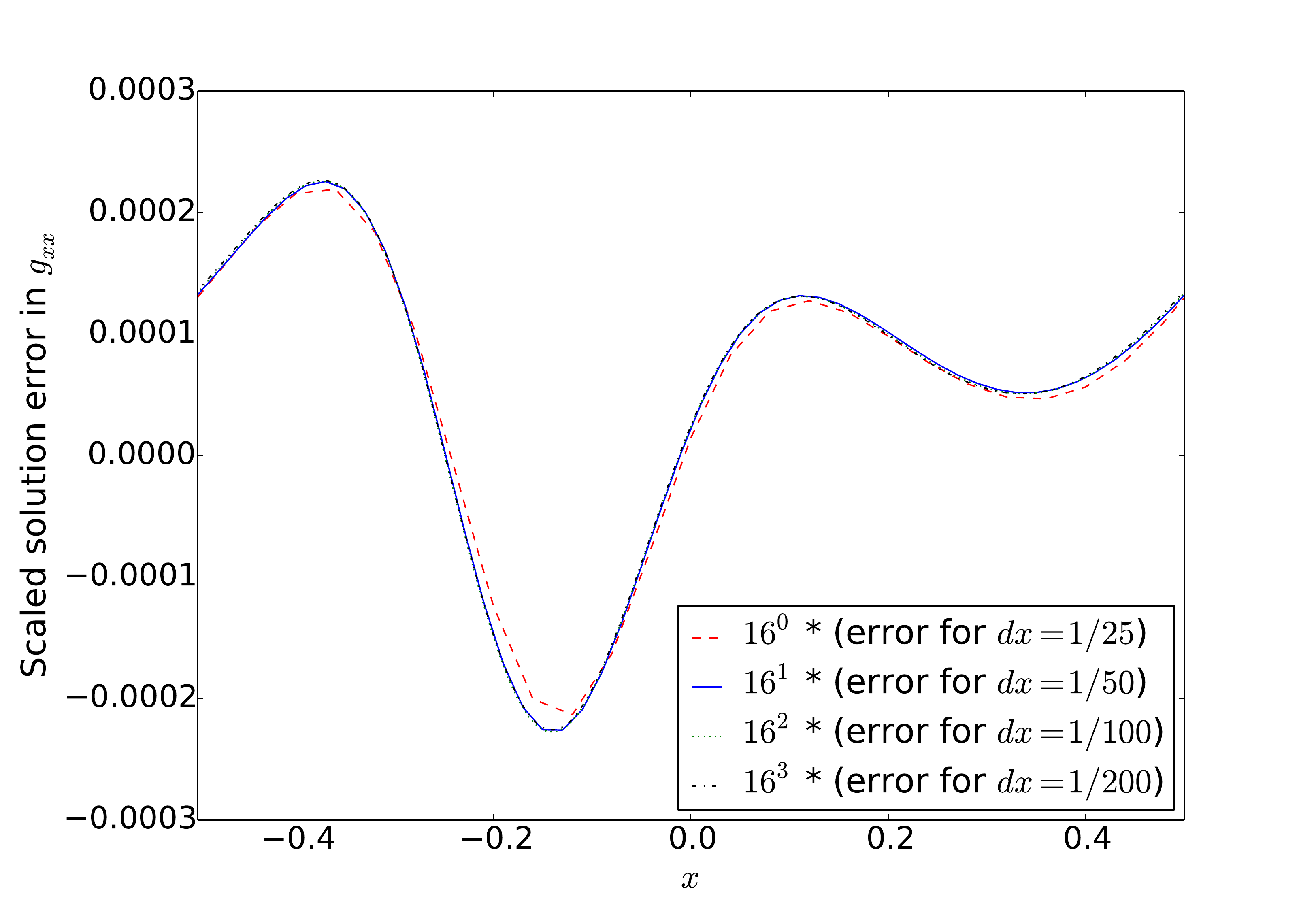}
\caption{\small{\textit{Scaled solution errors for the $g_{xx}$ component of the gauge wave metric after $2$ crossing times. The errors have been scaled with the resolution to highlight fourth order convergence.}}}
\label{fig:converge_gwave}
\end{figure}

In Figure \ref{fig:converge_gwave} we plot the solution errors for evolutions with resolution $dx=1/25\rho$ for $\rho=1,2,3$. The errors scale according to fourth order convergence.

%

\subsection{Wave equation: Gaussian pulse}
In the following test we show how our proposed algorithm handles artificial reflections that often arise when a propagating wave crosses a mesh refinement boundary. We are interested in waves propagating outward from the fine grid across mesh refinement boundaries into the coarser grid. We evolve the wave equation, with initial data given by a Gaussian pulse centered at the origin,
\begin{equation}
\phi(x,y,z, t=0) = A \exp(-x^{2}/\sigma^{2}) \qquad \Pi(x,y,z,t=0) = 0,
\end{equation}
with $\sigma=0.25$ and $A=1$. Although the wave propagates essentially in one dimension, we evolve it using the full 3D grid with periodic boundary conditions for the outer boundaries. The simulation domain covers $x\in [-4,4]$. The region $x\in[-1,1]$ is further refined by a factor of $r=2$.

\begin{figure}[h!]
\centering
\subfigure[]
          {\label{fig:sol}
            \includegraphics[height=6.5cm, width=0.48\textwidth]{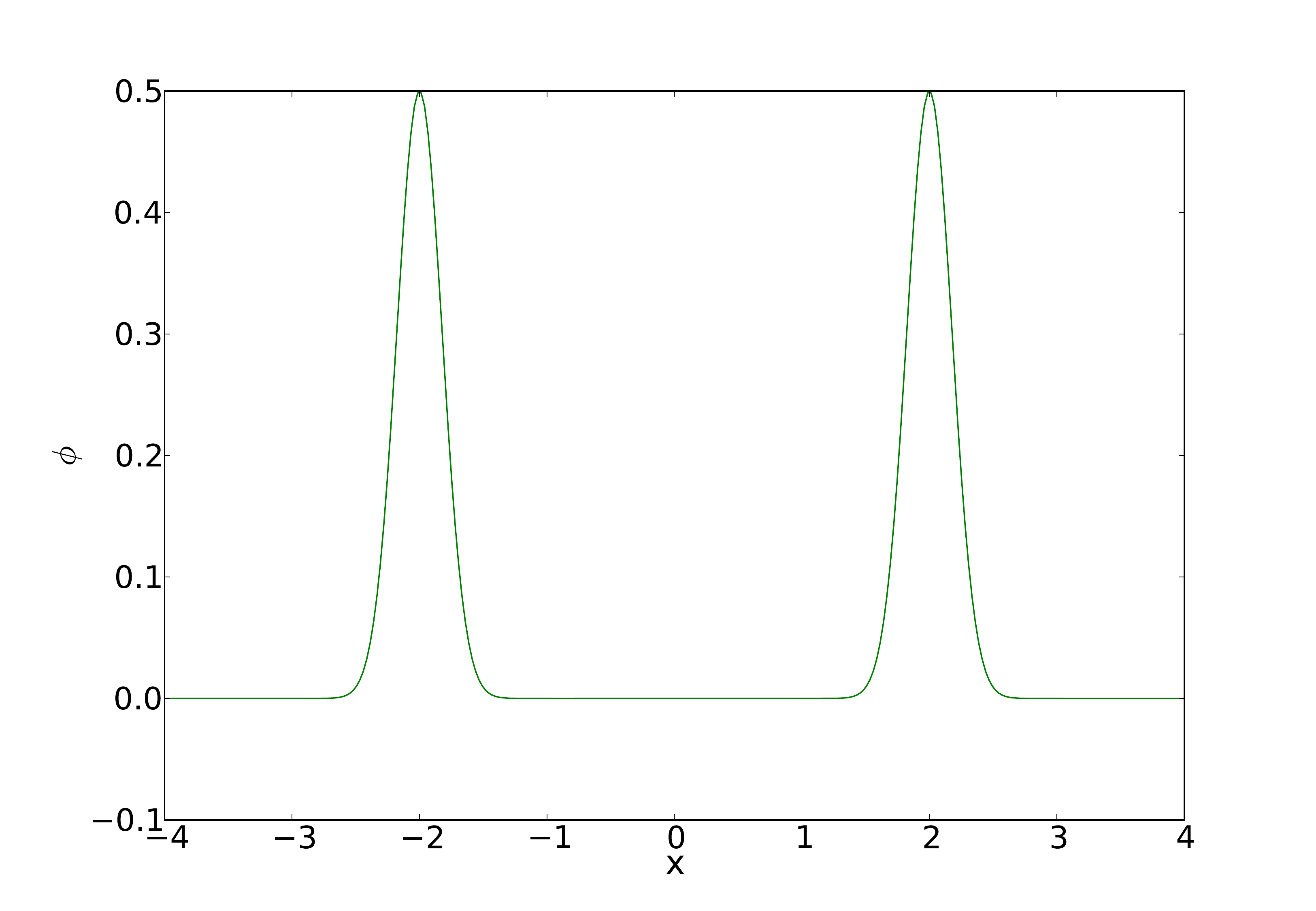}
          }
\subfigure[]
          {\label{fig:reflections}
            \includegraphics[height=6.5cm, width=0.48\textwidth]{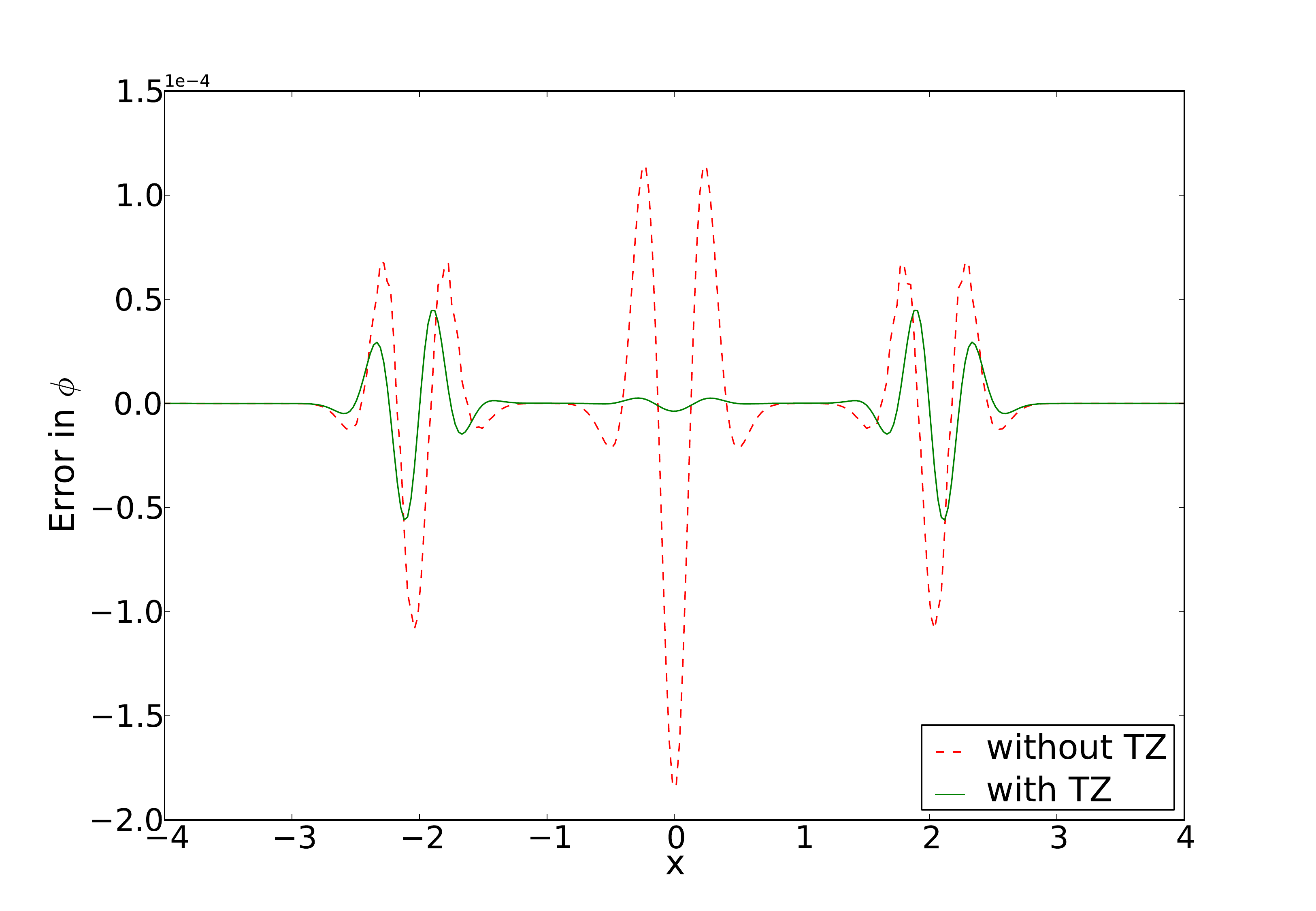}
          }
\caption{\small{\textit{Solution of the wave equation at $t=2s$. (a) A plot showing $\phi$. (b) Error in $\phi$ computed from the analytic solution for two runs with the same resolution, with and without a transition zone. Note the artificial reflection at $x=0$.}}}
\end{figure}

The solution is shown in Figure \ref{fig:sol}. The pulse starts initially at $x=0$ with amplitude one and produces two pulses each with amplitude $0.5$ traveling in opposite directions. In Figure
\ref{fig:reflections} we show the result after the pulses have crossed refinement boundaries at $x=\pm 1$. When each pulse crosses a refinement boundary, spurious reflections are generated. These
travel in a direction opposite that of the inducing pulse. When no transition zone is used, the spurious reflections reinforce at $x=0$ and can exceed the discretization error in amplitude. Employing
a transition zone significantly reduces these artificial reflections.

\subsection{Teukolsky Wave}
In this section we evolve the Einstein field equations in three space dimensions. As initial data, we use the Teukolsky solution for a quadrupole $l=2$, even parity $m=0$ waves \cite{PhysRevD.26.745}. The metric for the quadrupole modes is given by,
\begin{align}
ds^{2} &=-dt^{2}+(1+Af_{rr})dr^{2}+(2Bf_{r\theta})r\,dr\,d\theta + \left(2Bf_{r\phi} \right)r\,\sin\theta \,dr\,d\phi \\
& +\left(1+Cf^{(1)}_{\theta \theta}+Af^{(2)}_{\theta \theta}\right)r^{2}\,d\theta^{2} + [2(A-2C)f_{\theta \phi}] r^{2}\sin\theta \,d\theta \,d\phi \\
&+ \left(1+Cf^{(1)}_{\phi \phi}+Af^{(2)}_{\phi \phi}\right)r^{2}\,\sin^{2}\theta \,d\phi^{2}
\end{align}
The coefficients $A$, $B$ and $C$ are constructed via a generating function $F(x)$ and are given by,
\begin{align}
A &= 3\left[ \frac{F^{(2)}}{r^{3}} + \frac{3F^{(1)}}{r^{4}} + \frac{3F}{r^{5}} \right] \\
B &= -\left[ \frac{F^{(3)}}{r^{2}} +\frac{3F^{(2)}}{r^{3}} +\frac{6F^{(1)}}{r^{4}} + \frac{6F}{r^{5}}\right] \\
C &= \frac{1}{4}\left[\frac{F^{(4)}}{r} + \frac{2F^{(3)}}{r^{2}} + \frac{9F^{(2)}}{r^{3}} + \frac{21F^{(1)}}{r^{4}} + \frac{21F}{r^{5}}\right]
\end{align}
We take $F(x)$ to be a superposition of ingoing ($x=t+r$) and outgoing ($x=t-r$) waves,
\begin{equation}
F = F_{1}(t-r) + F_{2}(t+r)
\end{equation}
and,
\begin{equation}
F^{(n)} = \left[\frac{d^{n}\,F(x)}{dx^{n}}\right]_{x=t-r}+ (-1)^n \left[\frac{d^{n}\,F(x)}{dx^{n}}\right]_{x=t+r}
\end{equation}
where we have chosen the particular case $F_{1}(x)=-F_{2}(x)=\mathcal{A}e^{-x^{2}}$. The angular functions $f_{uv}$ for even parity $m=0$ modes are given by
\begin{align}
f_{rr} &= 2-3\sin^{2}\theta\\
f_{r\theta} &= -3\sin\theta \cos \theta\\
f_{r \phi} &= 0\\
f^{(1)}_{\theta \theta} &= 3\sin^{2}\theta\\
f^{(2)}_{\theta \theta} &= -1\\
f_{\theta \phi} &=  0\\
f^{(1)}_{\phi \phi} &= -f^{(1)}_{\theta \theta}\\ 
f^{(2)}_{\phi \phi} &= 3\sin^{2}\theta -1\\
\end{align}
We note that this represents time symmetric data and so $K_{ij}=0$ and $K=\gamma^{ij}K_{ij}=0$ at the initial slice $t=0$. We use an amplitude of $\mathcal{A}=10^{-6}$ to complete the specification of
initial data.

Because of the symmetries of the problem, we impose mirror symmetry boundary conditions along the planes $x=0$, $y=0$ and $z=0$. We thus evolve the Octant $[0,8]\times [0,8]\times[0,8]$ and use radiation boundary conditions at the outer boundary. We employ one refinement level and refine the cubic region $[0,4]\times[0,4]\times[0,4]$. 

\begin{figure}[!h]
\centering
\includegraphics[height=8.5cm,width=12cm]{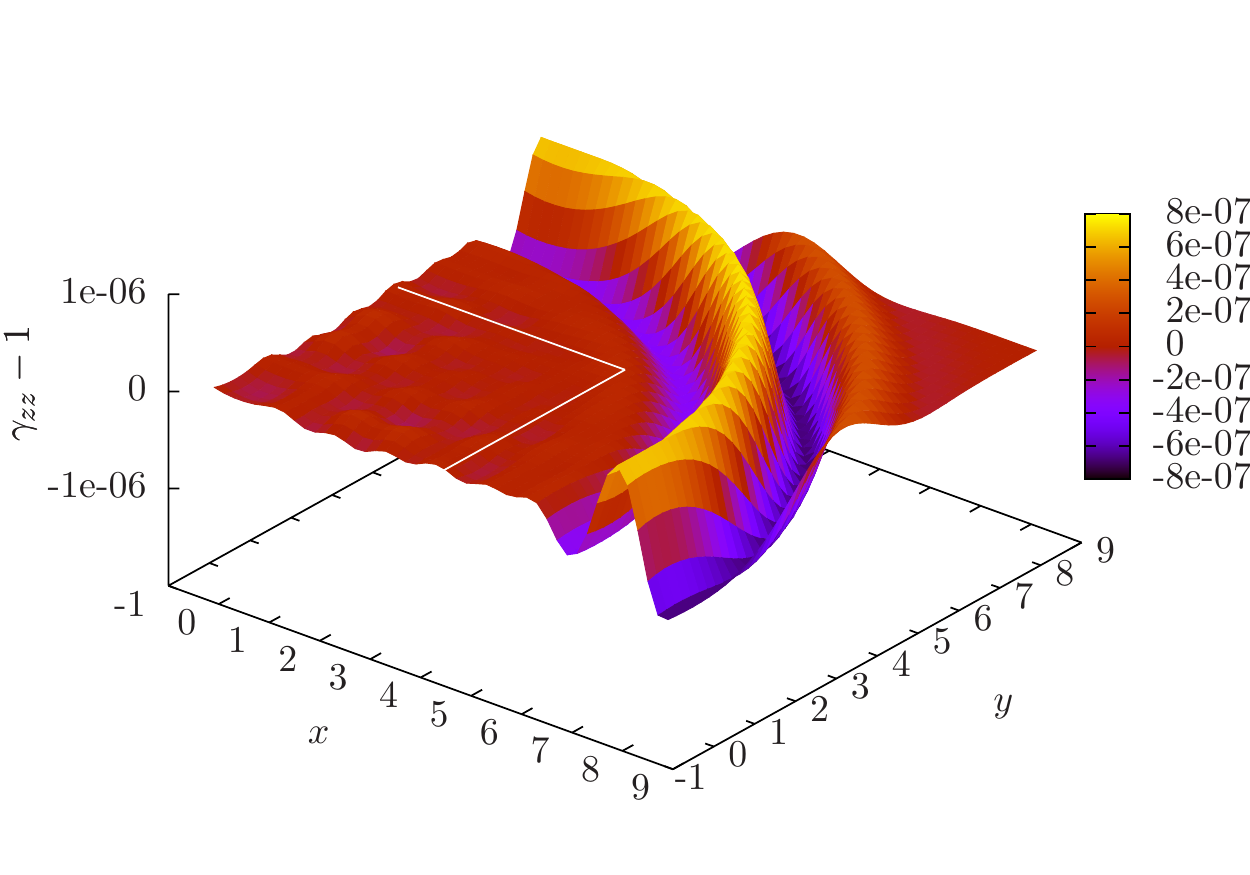}
\caption{\textit{\small{Evolution of the $\gamma_{zz}$ component of the metric along the $xy$ plane without a Transition zone. Note the spurious ripples in the refinement region. For ease of visualization, we mark the boundary of the refined grid with with white lines.}}}
\label{fig:ripples}
\end{figure}

\begin{figure}[!h]
\centering
\includegraphics[height=8.5cm,width=12cm]{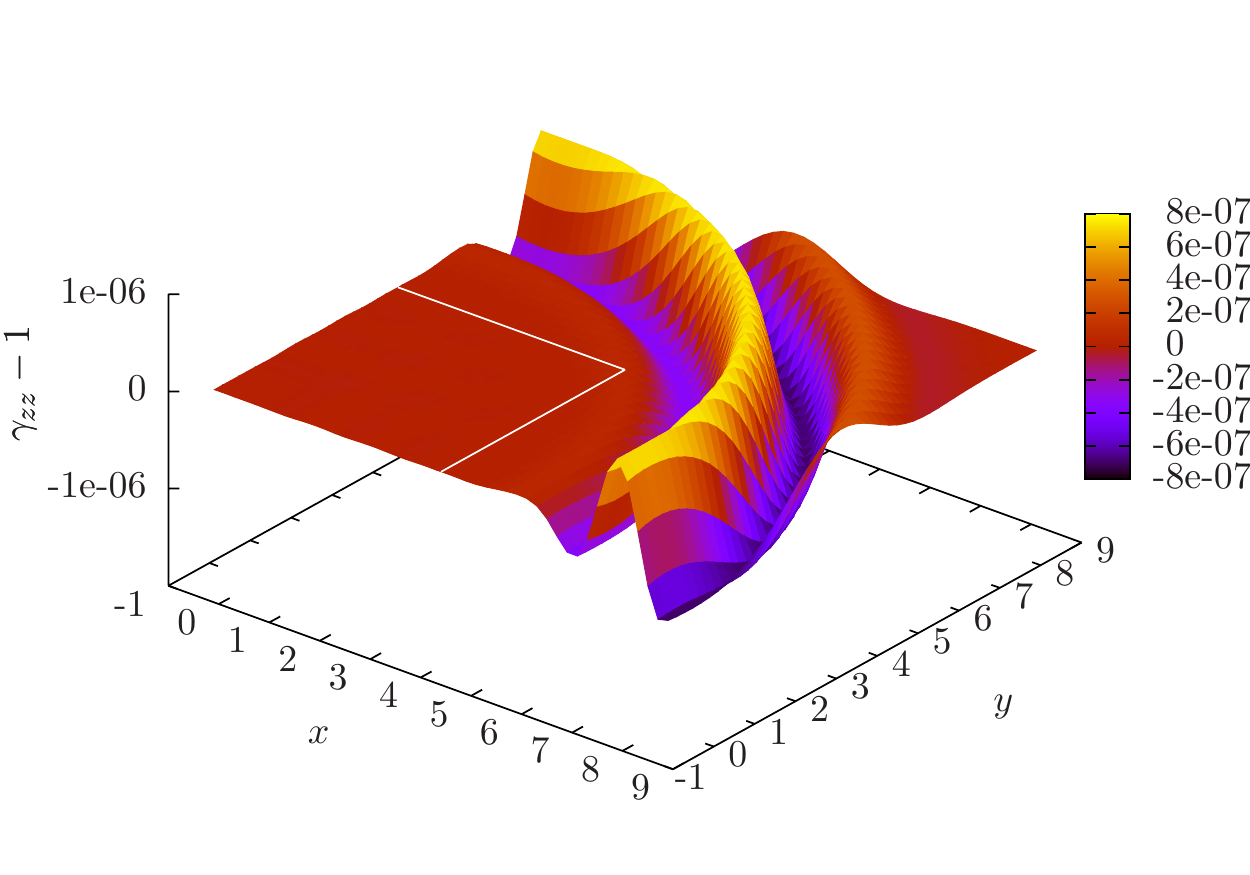}
\caption{\textit{\small{Evolution of the $\gamma_{zz}$ component of the metric along the $xy$ plane with a Transition zone. Note the absence of spurious ripples in the refinement region. For ease of visualization, we mark the boundary of the refined grid with with white lines. compare with Figure \ref{fig:ripples}.}}}
\label{fig:noripples}
\end{figure}

The wave propagates radially outward crossing mesh refinement boundaries along the $x=4$, $y=4$ and $z=4$ planes, eventually reaching the radiation boundary and leaving flat Minkowski spacetime.
Although the Teukolsky wave is a routine problem for testing numerical relativistic codes, it is especially challenging for a mesh refinement code. This is mainly because the refined region is
Cartesian, while the wave propagates spherically outward. As a result, the wavefront will not encounter the refinement boundaries at the same time. In Figures \ref{fig:ripples} we show the result at
$t=8$ for a run without a transition zone. Spurious ripples are generated when the wave initially hits the refinement boundary; the reflections continue to be generated until the wave has fully
crossed the refinement boundary. These ripples are reflected toward the origin as expected. In Figure \ref{fig:noripples}, we show a similar run using a transition zone. In this case, spurious
reflections are significantly minimized.

\section{Concluding Remarks}
\label{sec:concluding_remarks}

We have presented a fourth order mesh refinement scheme without the use of buffer zones. Our scheme also significantly minimizes spurious reflections off refinement boundaries that are caused by
differing levels of accuracy between two successive refinement levels. This is an important issue for the field of numerical relativity where the use of higher order finite differencing is becoming
increasingly common\cite{Pollney:2009yz,Diener:2005tn,Zlochower:2005bj}. For these higher order methods, the truncation error can become so small that the dominant error comes from spurious
reflections. Our method is not restricted to any formulation of the Einstein field equations. Indeed one can apply it to any hyperbolic system of partial differential equations.

Because we are not using a buffer zone, our method requires a total of six points on each boundary, irrespective of the time integration method used. This differs from employing buffer zones in that,
for a fourth order accurate Runge Kutta algorithm, along with fourth order finite differencing involving lop-sided advection stencils, $12$ points are needed along each boundary (the situation could
be worse for higher order finite differencing) \cite{Husa:2007hp,Bruegmann:2006at}. This is a significant saving in memory usage, especially in three space dimensions where the buffer zone can be a
significant part of the grid. In addition, the blending operation we employ to fill the transition zone is cheaper than having to repopulate the entire buffer zone after every time step.

We also note that the use of a transition zone is computationally cheaper than the sponge boundary method since one has to populate the sponge boundary at every intermediate Runge Kutta step, while
the transition zone is only populated at the end of the time step. Moreover, there is often a level of experimentation required to determine how large a sponge zone one should use. Although one can
have the transition zone as large as desired, we have found satisfactory results with a size that spans only three fine grid points.

The implementation described here uses Runge Kutta dense output formulas to interpolate in time, which avoids any potential issues with polynomial interpolation. In particular, because one does not have to couple fine grid solutions to the solution history of coarser grids, fine grids can be immediately initialized along with the base grid. This is especially attractive since fine grids can be initialized to the same accuracy as the base grid, by using the same initialization routine as the base grid. This method of time interpolation was also used in \cite{mccorquodale2011high} for conservation laws and \cite{chiltonThesis} for solving Maxwell's equations.

It would be interesting to investigate the efficiency of the transition zone implementation in an adaptive context or whether it would minimize spurious reflections that are a result of shock waves
crossing refinement boundaries. There is also the question of how it would affect conservation along interface boundaries in the context of conservation laws. 
These and other issues involving black hole spacetimes are a subject of further study.

\ack
I thank Denis Pollney for discussions and comments.
Some of the computations were performed using facilities provided by the University of Cape Town's ICTS High Performance Computing team.

\section{Appendix}
\subsection{Dispersion relation}
\label{sec:dispersion}
In this section we calculate the dispersion relation and phase velocity resulting in discretizing the wave equation with a fourth order stencil along with Runge Kutta time marching. We follow the approach (and notation) of \cite{2004JCoPh.193..398C} where a similar calculation was given using second order finite differences and iterative Crank-Nicholson time marching. 

The wave equation \ref{eq:wave_eq2} can be written in matrix form as
\begin{equation}
\label{eq:waveMatrix}
V_{t} = 
\left(
      \begin{array}{cc}
         0 & 1 \\
        \pp_{xx} & 0
       \end{array}
 \right) V
\end{equation}
where we have defined the vector $V$ as
\begin{equation}
V = 
\left(
      \begin{array}{c}
         \phi  \\
         \Pi
       \end{array}
 \right) 
\end{equation}
With this identification, we denote by $V^{n}_{\phantom{j}j}$ the solution at time step $n$ and grid point $j$.
%
The second derivative operator $\pp_{xx}$ appearing in \ref{eq:waveMatrix} is given by the stencil,
\begin{align}
  \pp_{xx} V^{n}_{i} = \frac{- V^{n}_{\phantom{n}i+2} + 16 V^{n}_{\phantom{n}i+1} -30 V^{n}_{\phantom{n}i} + 16 V^{n}_{\phantom{n}i-1} - V^{n}_{\phantom{n}i-2}}{12 dx^2}.
\end{align}

For our analysis, we consider plane wave solutions for $V$, of the form,
\begin{equation}
\label{eq:planeWaves}
V^{n}_{j} = W e^{i\omega n dt} e^{-ikj dx},
\end{equation}
for some constant vector $W$. Using the classical fourth order Runge Kutta scheme to advance \ref{eq:waveMatrix} in time, results in the update rule,
\begin{equation}
V^{n+1}_{\phantom{n+1}j} = M V^{n}_{\phantom{n}j}.
\end{equation}
Plugging in \ref{eq:planeWaves} to the above rule results in the relation,
\begin{equation}
\label{eq:eigenvalueEquation}
e^{i \omega dt} W = 
\left(
      \begin{array}{cc}
        1-2\Lambda^{2}+\frac{2}{3}\Lambda^{4} & dt\left(1-\frac{2}{3}  \Lambda^{2} \right) \\
        -\dfrac{4 \Lambda^{2} (3-2\Lambda^{2})}{3dt} & 1-2\Lambda^{2}+\frac{2}{3}\Lambda^{4}
       \end{array}
 \right) W
\end{equation}
where we have defined $\Lambda$ as,
\begin{equation}
\Lambda = \frac{dt}{dx} \sqrt{\frac{4}{3} \sin^{2}\left(\frac{k dx}{2}\right)-\frac{1}{12} \sin^{2}\left(k dx\right)}
\end{equation}
The system \ref{eq:eigenvalueEquation} represents an eigenvalue problem. In particular, $W$ is an eigenvector corresponding to the eigenvalue $e^{i\omega dt}$ for the matrix in \ref{eq:eigenvalueEquation}. Further analysis shows the eigenvalues to be the pair,
\begin{equation}
\label{eq:eigenvalues}
e^{i \omega dt}  = 1-2\Lambda^{2}+\frac{2}{3}\Lambda^{4} \pm 2i\Lambda\left(1-\frac{2}{3}\Lambda^{2}\right)
\end{equation}
This expression represents the dispersion relation, relating the frequency $\omega$ with the wave number $k$. For completeness we calculate the phase velocity,
$v_{p}(\lambda) = \xi/k$ for $\xi=\mathrm{Re}\;(\omega)$.
From \ref{eq:eigenvalues} we get,
\begin{equation}
\xi dt = \arcsin\left( \frac{2\Lambda (3-2\Lambda^{2})}{ \sqrt{9-4\Lambda^{6}(2-\Lambda^{2})}} \right)
\end{equation}
Therefore,
\begin{eqnarray}
v_{p}(\lambda) &=& \frac{\lambda}{2\pi dt} \arcsin\left( \frac{2\Lambda (3-2\Lambda^{2})}{ \sqrt{9-4\Lambda^{6}(2-\Lambda^{2})}} \right)
\end{eqnarray}
\section*{References}
\bibliographystyle{plain}
\bibliography{ref}
\end{document}